# Monte Carlo Calculations of the Extraction of Scintillation Light from Cryogenic *N*-type GaAs

Stephen Derenzo
*Retired Affiliate*
Lawrence Berkeley National Laboratory

sederenzo@lbl.gov

Abstract

The high scintillation luminosity of *n*-type GaAs at 10K is surprising because (1) with a refractive index of about 3.5, escape is inhibited by total internal reflection and (2) narrow-beam experiments at 90K report infrared absorption coefficients of several per cm. This paper presents Monte Carlo calculations showing that the high luminosity at 10K can be explained if (1) narrow-beam absorption is almost all optical scattering and (2) the absolute absorption coefficient is below 0.1 per cm. Sixteen surface reflector configurations are simulated for a range of internal scattering and absolute absorption coefficients, and these can guide the design of cryogenic scintillating GaAs targets for the direct detection of dark matter. The discussion section presents a possible infrared scattering mechanism based on the metallic nature of *n*-type GaAs. Appendix A describes the Monte Carlo program steps in detail. Appendix B shows how narrow-beam and integrating sphere experiments can measure the cryogenic optical scattering and absolute absorption coefficients.

1. Introduction

In previous work, we determined that *n*-type GaAs doped with Si and B is a luminous scintillator at 10K [1, 2] and explored the dependence of its scintillation properties on the concentrations of Si and B [3]. These measurements were done with 40 to 65 kVp X-rays that deposited ionization energy hundreds of μm within the crystals. The cryogenic band gap is 1.52 eV (816 nm) and the four principal emission bands peak at 860, 930, 1070, and 1335 nm. Externally measured luminosities as high as 110 photons/keV from bare GaAs crystals with external reflectors are surprising because with a refractive index of 3.5 [4] escape is inhibited by total internal reflection and narrow-beam experiments have shown significant absorption. The luminosity of 110 photons/keV corresponds to an exit fraction of 50%, assuming that every ionization hole recombines radiatively with a free (conduction band) electron. The theoretical limit is 220 photons/keV, based on the pair creation energy of 4.55 ± 0.02 eV at 0K [5]. Any ionization holes that recombine non-radiatively would decrease the number of photons/keV and make the calculated exit fraction >50%.

Osamura et al. measured narrow-beam infrared absorption coefficients of *n*-type GaAs as a function of the free electron concentration. At 90K the absorption coefficient for 2,500 nm photons varied linearly from 0.1/cm to 10/cm for free electron concentrations from $10^{16}$ to $10^{18}$/cm$^3$ [6]. Spitzer et al. measured narrow-beam absorption coefficients as a function of temperature. For 1,500 nm photons and a free electron concentration of 4.7 x $10^{17}$/cm$^3$ the coefficients were 2.5/cm, 3.1/cm, and 3.8/cm at 100K, 297K and 443K, respectively [7]. Undoped GaAs has much less narrow-beam absorption [8] and is used as an optical filter that transmits in the infrared. In this paper the term narrow-beam absorption includes both optical scattering and absolute absorption. Photons are absolutely absorbed when their energy is entirely transferred to the lattice, such as by intraband absorption [9]. The terms internal scattering and internal absolute absorption are used for the interaction of photons with free electrons in the volume of the crystal.

This paper is organized as follows. Section 2 describes 16 different surface reflector configurations, and summarizes the Monte Carlo program steps and statistical uncertainties. Section 3 tabulates the fractions of scintillation photons that exit, are trapped, or are absorbed for the different surface treatments, thicknesses, and for different values of the scattering and absolute absorption coefficients. The average path length in the crystal before exit or absorption is also tabulated. Section 4 discusses the merits of cryogenic $n$-type GaAs for the direct detection of dark matter and suggests a possible scattering mechanism. Section 5 lists the conclusions. Appendix A describes the Monte Carlo program steps in detail. Appendix B describes how narrow-beam and integrating sphere experiments can measure the optical scattering and absolute absorption coefficients.

2. Computational Methods

The GaAs crystal is modeled as a rectangular solid with a 1 cm x 1 cm top exit face and opposing bottom face, and thicknesses from 0.03 cm to 1 cm. The refractive index is chosen to be 3.5 [4], which is close to the average over the wavelengths of the scintillation emissions [2].

Four surface treatments are considered as follows:
- PM designates a polished surface that is coated with a metal (such as gold) to form an optically bonded mirror that reflects 98% of the incident photons back into the crystal and absorbs 2%.
- PD designates a polished surface that has a diffuse Lambertian reflector (e.g. white paint) optically bonded to its surface. Photons reaching the surface are reflected with 98% probability into a random hemispherical distribution back into the crystal or are absorbed with 2% probability. An alternative design is a rough crystal surface that randomizes the internal reflection angles but is also translucent so an external diffuse reflector is needed to redirect photons back to the crystal.
- PEM designates a bare polished surface and an external mirror reflector. Photons that are refracted into the narrow space between the surface and the mirror are reflected with 98% probability back toward the crystal or are absorbed with 2% probability.
- PED designates a bare polished surface and an external diffuse Lambertian reflector. Photons that are refracted into the narrow space between the surface and the diffuse reflector are reflected with 98% probability into a random hemispherical distribution back toward the crystal or are absorbed with 2% probability.

The bottom face can be PM, PD, PEM or PED. The four side faces all have the same treatment, which can be PM, PD, PEM or PED. This results in the 16 combinations listed in the first five tables of Section 3. Reflectances of 98% for both the diffuse and mirror reflectors were chosen because they are values commonly achieved for the interior surfaces of integrating spheres and the mirrors of infrared telescopes.

In each of the cases listed in the tables below, 1,000,000 photons were selected from a uniformly random spatial distribution throughout the crystal and with an isotropically random distribution in angle. Individual photons were tracked until they exited the top face, were internally trapped or absorbed, or were absorbed by the reflectors (see Appendix A for details). Each photon has a weight of one until it exits or is absorbed (rather than splitting into multiple photons with fractional weights at decision points). For the 1,000,000 photons used in each case, exemplar absorption fractions of 4%, 10%, 50% and 90% have standard deviations of 0.019%, 0.030%, 0.051%, and 0.030%, respectively.

Section 3. Results

Because of total internal reflection, only photons within an escape cone can exit a polished GaAs surface. For isotropic photons distributed throughout a polished GaAs cube, 2.09% of the photons escape from each of the six faces and the remaining 87.46% are trapped by total internal reflection. This is very different from conventional scintillators that have lower refractive indexes and much larger escape cone solid angles.

Table 1 lists the fractions of photons that exit the top face, are absorbed on the bottom or side faces, or are trapped for the case of a 1 cm cube with no internal scattering or absolute absorption. Combinations that have only mirror or total internal reflection faces (PM, PEM, or PED) and no PD faces have exit fractions around 4%. Without a PD surface to scatter photons into the escape cone, 2% exit the top face directly and 2% are reflected into the escape cone from the bottom face. The exit fractions are much higher if one or more diffuse (PD) faces are involved. Combinations 2, 10 and 14 have exit fractions near 53% and diffuse side reflectors (PD) that scatter photons into the top face exit cone directly or via bottom face mirror reflections (PM) or bottom face total internal reflections (PEM or PED). Combination 6 with PD bottom and side faces has an exit fraction of 41.5%. Combinations 7 and 8 have exit fractions around 36%, diffuse bottom face reflectors (PD), and total internal reflection sides (PEM or PED).

Table 1. Results for 1,000,000 photons generated within a 1 cm GaAs cube, no internal scattering or absorption.

| Surface combi-nation | Bottom face | 4 side faces | Exit top face | Absorbed at bottom face | Absorbed at 4 side faces | Trapped by total internal reflection | Average path length (cm) |
|---|---|---|---|---|---|---|---|
| 1 | PM | PM | 4.02% | 19.30% | 76.67% | 0.00% | 38.1 |
| 2 | PM | PD | 54.08% | 10.08% | 35.83% | 0.00% | 17.2 |
| 3 | PM | PEM | 4.08% | 87.84% | 8.08% | 0.00% | 837 |
| 4 | PM | PED | 4.10% | 87.76% | 8.14% | 0.00% | 907 |
| 5 | PD | PM | 13.35% | 8.75% | 77.91% | 0.00% | 33.1 |
| 6 | PD | PD | 41.64% | 13.14% | 45.21% | 0.00% | 19.9 |
| 7 | PD | PEM | 36.69% | 25.52% | 37.79% | 0.00% | 303 |
| 8 | PD | PED | 36.69% | 25.52% | 37.79% | 0.00% | 302 |
| 9 | PEM | PM | 3.92% | 0.24% | 95.84% | 0.00% | 50.2 |
| 10 | PEM | PD | 53.24% | 3.58% | 43.18% | 0.00% | 20.5 |
| 11 | PEM | PEM | 3.93% | 0.24% | 6.97% | 88.86% | trapped |
| 12 | PEM | PED | 3.95% | 0.23% | 6.91% | 88.91% | trapped |
| 13 | PED | PM | 3.85% | 0.23% | 95.92% | 0.00% | 50.5 |
| 14 | PED | PD | 53.28% | 3.60% | 43.12% | 0.00% | 20.5 |
| 15 | PED | PEM | 3.93% | 0.24% | 6.91% | 88.92% | trapped |
| 16 | PED | PED | 3.92% | 0.24% | 6.92% | 88.92% | trapped |

Table 2 lists the surface combinations of Table 1 with the addition of an internal absolute absorption coefficient of 0.1/cm. All exit fractions are significantly reduced. The average path lengths in the crystal are reduced because many photons are absorbed before they can exit. Table 3 lists the surface combinations of Table 1 with an internal absolute absorption coefficient of 1/cm. Exit fractions are reduced to values between 1.2% and 2%, and the average path lengths are reduced to about 1 cm.

Table 2. Results for 1,000,000 photons generated within a 1 cm GaAs cube, no internal scattering, and absolute absorption coefficient 0.1/cm.

| Surface combi-nation | Bottom face | 4 side faces | Exit top face | Absorbed at bottom face | Absorbed at 4 side faces | Absolutely absorbed internally | Average path length (cm) |
|---|---|---|---|---|---|---|---|
| 1 | PM | PM | 3.37% | 3.78% | 15.84% | 77.01% | 7.7 |
| 2 | PM | PD | 19.06% | 3.75% | 13.38% | 63.80% | 6.4 |
| 3 | PM | PEM | 3.41% | 4.38% | 2.84% | 89.37% | 8.9 |
| 4 | PM | PED | 3.39% | 4.40% | 2.85% | 89.36% | 8.9 |
| 5 | PD | PM | 4.74% | 2.74% | 17.02% | 75.49% | 7.6 |
| 6 | PD | PD | 13.79% | 4.33% | 15.12% | 66.77% | 6.7 |
| 7 | PD | PEM | 5.44% | 3.28% | 4.61% | 86.67% | 8.7 |
| 8 | PD | PED | 5.43% | 3.28% | 4.61% | 86.68% | 8.7 |
| 9 | PEM | PM | 3.23% | 0.20% | 16.41% | 80.16% | 8.0 |
| 10 | PEM | PD | 17.04% | 1.16% | 14.27% | 67.54% | 6.8 |
| 11 | PEM | PEM | 3.21% | 0.20% | 2.86% | 93.73% | 9.4 |
| 12 | PEM | PED | 3.25% | 0.19% | 2.86% | 93.70% | 9.4 |
| 13 | PED | PM | 3.23% | 0.19% | 16.42% | 80.16% | 8.0 |
| 14 | PED | PD | 17.11% | 1.17% | 14.31% | 67.41% | 6.8 |
| 15 | PED | PEM | 3.24% | 0.20% | 2.85% | 93.71% | 9.3 |
| 16 | PED | PED | 3.24% | 0.19% | 2.86% | 93.71% | 9.4 |

Table 3. Results for 1,000,000 photons generated within a 1 cm GaAs cube, no internal scattering, and internal absolute absorption coefficient 1/cm.

| Surface combi-nation | Bottom face | 4 side faces | Exit top face | Absorbed at bottom face | Absorbed at 4 side faces | Absolutely absorbed internally | Average path length (cm) |
|---|---|---|---|---|---|---|---|
| 1 | PM | PM | 1.24% | 0.48% | 1.95% | 96.33% | 0.96 |
| 2 | PM | PD | 2.12% | 0.52% | 2.04% | 95.32% | 0.95 |
| 3 | PM | PEM | 1.25% | 0.49% | 0.45% | 97.81% | 0.98 |
| 4 | PM | PED | 1.25% | 0.49% | 0.45% | 97.81% | 0.98 |
| 5 | PD | PM | 1.10% | 0.47% | 2.03% | 96.40% | 0.96 |
| 6 | PD | PD | 1.82% | 0.57% | 2.14% | 95.47% | 0.96 |
| 7 | PD | PEM | 1.12% | 0.49% | 0.51% | 97.89% | 0.98 |
| 8 | PD | PED | 1.13% | 0.48% | 0.52% | 97.86% | 0.98 |
| 9 | PEM | PM | 1.21% | 0.09% | 1.96% | 96.74% | 0.97 |
| 10 | PEM | PD | 2.01% | 0.16% | 2.03% | 95.80% | 0.96 |
| 11 | PEM | PEM | 1.21% | 0.09% | 0.45% | 98.26% | 0.98 |
| 12 | PEM | PED | 1.22% | 0.09% | 0.45% | 98.24% | 0.98 |
| 13 | PED | PM | 1.21% | 0.09% | 1.96% | 96.74% | 0.97 |
| 14 | PED | PD | 2.01% | 0.15% | 2.03% | 95.81% | 0.96 |
| 15 | PED | PEM | 1.22% | 0.09% | 0.45% | 98.25% | 0.98 |
| 16 | PED | PED | 1.22% | 0.08% | 0.45% | 98.24% | 0.98 |

Table 4 lists the surface combinations of Table 1 with an internal scattering coefficient of 1/cm and no internal absolute absorption. The exit fraction is increased for all 16 surface treatments. The highest exit fractions occur for combinations 11, 12, 15, and 16 (PEM or PED only) where many of the photons that were trapped by total internal reflection (88.9% in Table 1) now can scatter into the exit cone, resulting in an increase of the exit fractions from 4% to over 69%. The scattering provides an explanation for the high luminosities of the bare *n*-type GaAs crystals with external reflectors reported in refs [1, 3].

Table 4. Results for 1,000,000 photons generated within a 1 cm GaAs cube, internal scattering coefficient 1/cm, and no internal absolute absorption.

| Surface combi-nation | Bottom face | 4 side faces | Exit top face | Absorbed at bottom face | Absorbed at 4 side faces | Average path length (cm) |
|---|---|---|---|---|---|---|
| 1 | PM | PM | 33.73% | 13.16% | 53.11% | 26.2 |
| 2 | PM | PD | 44.70% | 11.11% | 44.19% | 20.2 |
| 3 | PM | PEM | 57.09% | 22.37% | 20.54% | 44.9 |
| 4 | PM | PED | 57.08% | 22.35% | 20.57% | 44.8 |
| 5 | PD | PM | 30.58% | 13.71% | 55.71% | 26.7 |
| 6 | PD | PD | 39.09% | 13.24% | 47.67% | 20.8 |
| 7 | PD | PEM | 52.15% | 23.63% | 24.23% | 45.7 |
| 8 | PD | PED | 52.13% | 23.65% | 24.22% | 45.6 |
| 9 | PEM | PM | 36.99% | 2.67% | 60.34% | 29.7 |
| 10 | PEM | PD | 47.02% | 3.62% | 49.36% | 22.5 |
| 11 | PEM | PEM | 69.34% | 5.01% | 25.66% | 56.2 |
| 12 | PEM | PED | 69.33% | 5.06% | 25.60% | 56.2 |
| 13 | PED | PM | 37.16% | 2.71% | 60.13% | 29.6 |
| 14 | PED | PD | 46.87% | 3.59% | 49.54% | 22.5 |
| 15 | PED | PEM | 69.23% | 5.07% | 25.71% | 56.2 |
| 16 | PED | PED | 69.30% | 5.05% | 25.65% | 56.1 |

Table 5 lists the surface combinations of Table 1 for an internal scattering coefficient of 1/cm and an absolute absorption coefficient of 0.1/cm. Most of the photons are internally absorbed before they can scatter into the escape cone, resulting in low exit fractions for all combinations.

Table 5. Results for 1,000,000 photons generated within a 1 cm GaAs cube, internal scattering coefficient 1/cm, and absolute absorption coefficient 0.1/cm.

| Surface combi-nation | Bottom face | 4 side faces | Exit top face | Absorbed at bottom face | Absorbed at 4 side faces | Absolutely absorbed internally | Average path length (cm) |
|---|---|---|---|---|---|---|---|
| 1 | PM | PM | 9.44% | 3.63% | 14.65% | 72.29% | 7.2 |
| 2 | PM | PD | 14.55% | 3.68% | 14.66% | 67.11% | 6.7 |
| 3 | PM | PEM | 10.54% | 4.11% | 3.80% | 81.56% | 8.2 |
| 4 | PM | PED | 10.56% | 4.11% | 3.75% | 81.58% | 8.2 |
| 5 | PD | PM | 8.47% | 3.74% | 15.16% | 72.63% | 7.3 |
| 6 | PD | PD | 12.67% | 4.22% | 15.39% | 67.72% | 6.8 |
| 7 | PD | PEM | 9.44% | 4.24% | 4.34% | 81.98% | 8.2 |
| 8 | PD | PED | 9.51% | 4.29% | 4.33% | 81.87% | 8.2 |
| 9 | PEM | PM | 9.49% | 0.70% | 15.16% | 74.65% | 7.5 |
| 10 | PEM | PD | 14.29% | 1.09% | 15.21% | 69.41% | 6.9 |
| 11 | PEM | PEM | 10.66% | 0.80% | 3.87% | 84.67% | 8.5 |
| 12 | PEM | PED | 10.68% | 0.79% | 3.84% | 84.70% | 8.5 |
| 13 | PED | PM | 9.53% | 0.70% | 15.10% | 74.67% | 7.5 |
| 14 | PED | PD | 14.25% | 1.11% | 15.20% | 69.44% | 6.9 |
| 15 | PED | PEM | 10.63% | 0.81% | 3.86% | 84.70% | 8.5 |
| 16 | PED | PED | 10.63% | 0.78% | 3.90% | 84.69% | 8.5 |

Table 6 lists the fraction of photons that are trapped, exit, or are absorbed for the case of a 1 cm cube with external mirrors (PEM) on the bottom and side faces, and with different values of internal scattering and absolute absorption coefficients. With no internal scattering (rows 1-4) the exit fraction is only about 4%, which is the sum of the direct top face exit fraction and reflections from the bottom face. The addition of internal absorption (rows 2-4) does not affect the exit fraction and only serves to absolutely absorb photons that would otherwise be trapped by total internal reflection. An absolute absorption coefficient of 0.001/cm corresponds to an absorption length of 1000 cm, which is close to the average path length of 877 cm (row 2). With no absolute absorption a small amount of scattering (0.01/cm in row 5) dramatically increases the exit fraction by scattering photons into the top face exit cone. Adding absolute absorption in rows 6-8 shows a dramatic decrease in the exit fraction by absorbing photons before they can be scattered into the top face exit cone. Increasing the scattering coefficient to 0.1/cm repeats the same pattern. Increasing the scattering coefficient to 1/cm and 10/cm (rows 13-20) results in about the same exit fractions, showing the insensitivity to scattering in this range. An absolute absorption coefficient of 0.1/cm (rows 3, 8, 12, 16, 20) dramatically limits the exit fraction. In summary, a high exit fraction is associated with scattering coefficients at or above 0.1/cm and absolute absorption coefficients below 0.1/cm.

Table 6. Results for 1,000,000 photons generated within a 1 cm GaAs cube with PEM bottom and side faces for different internal scattering and absolute absorption coefficients.

| Row | Scattering coefficient (cm$^{-1}$) | Absolute absorption coefficient (cm$^{-1}$) | Exit top face | Absorbed at bottom face | Absorbed at 4 side faces | Trapped by total internal reflection | Absolutely absorbed internally | Average path length (cm) |
|---|---|---|---|---|---|---|---|---|
| 1 | 0 | 0 | 3.95% | 0.24% | 6.91% | 88.90% | 0.00% | trapped |
| 2 | 0 | 0.001 | 3.91% | 0.24% | 8.16% | 0.00% | 87.70% | 876.9 |
| 3 | 0 | 0.01 | 3.88% | 0.23% | 6.95% | 0.00% | 88.94% | 88.9 |
| 4 | 0 | 0.1 | 3.28% | 0.20% | 2.87% | 0.00% | 93.65% | 9.4 |
| 5 | 0.01 | 0 | 35.00% | 2.11% | 62.88% | 0.00% | 0.00% | 804.9 |
| 6 | 0.01 | 0.001 | 20.21% | 1.19% | 35.92% | 0.00% | 42.68% | 426.3 |
| 7 | 0.01 | 0.01 | 6.85% | 0.42% | 10.90% | 0.00% | 81.83% | 81.8 |
| 8 | 0.01 | 0.1 | 3.52% | 0.21% | 2.94% | 0.00% | 93.33% | 9.3 |
| 9 | 0.1 | 0 | 51.28% | 3.19% | 45.53% | 0.00% | 0.00% | 148.0 |
| 10 | 0.1 | 0.001 | 44.93% | 2.76% | 39.46% | 0.00% | 12.85% | 128.4 |
| 11 | 0.1 | 0.01 | 21.92% | 1.37% | 18.48% | 0.00% | 58.23% | 58.3 |
| 12 | 0.1 | 0.1 | 5.26% | 0.34% | 3.37% | 0.00% | 91.03% | 9.1 |
| 13 | 1 | 0 | 69.30% | 5.08% | 25.63% | 0.00% | 0.00% | 56.2 |
| 14 | 1 | 0.001 | 65.66% | 4.78% | 24.25% | 0.00% | 5.31% | 53.1 |
| 15 | 1 | 0.01 | 44.41% | 3.22% | 16.39% | 0.00% | 35.98% | 35.9 |
| 16 | 1 | 0.1 | 10.61% | 0.79% | 3.91% | 0.00% | 84.69% | 8.5 |
| 17 | 10 | 0 | 67.98% | 6.66% | 25.36% | 0.00% | 0.00% | 53.5 |
| 18 | 10 | 0.001 | 64.61% | 6.34% | 24.00% | 0.00% | 5.04% | 50.9 |
| 19 | 10 | 0.01 | 44.33% | 4.36% | 16.45% | 0.00% | 34.86% | 34.9 |
| 20 | 10 | 0.1 | 10.85% | 1.03% | 3.96% | 0.00% | 84.16% | 8.4 |

Table 7 lists the exit and absorption fractions for the thin (0.03 to 0.3 cm thick) GaAs crystals that were reported in [3]. It shows that thinner crystals have somewhat higher exit fractions but the conclusions are the same as for the 1 cm cubes, namely that a high exit fraction requires a scattering coefficient at or above 0.1/cm and an absolute absorption coefficient below 0.1/cm.

Table 7. Results for 1,000,000 photons generated in a rectangular GaAs crystal with PEM bottom and side facts for different thicknesses, and for different internal scattering and absolute absorption coefficients.

| Thickness (cm) | Scattering coefficient (cm$^{-1}$) | Absorption coefficient (cm$^{-1}$) | Exit top face | Absorbed at bottom face | Absorbed at 4 side faces | Absolutely absorbed internally | Average path length (cm) |
|---|---|---|---|---|---|---|---|
| 1 | 0.1 | 0 | 51.41% | 3.17% | 45.42% | 0.00% | 148.4 |
| 0.3 | 0.1 | 0 | 54.44% | 3.30% | 42.25% | 0.00% | 137.4 |
| 0.1 | 0.1 | 0 | 55.31% | 3.36% | 41.33% | 0.00% | 133.7 |
| 0.03 | 0.1 | 0 | 55.68% | 3.35% | 40.96% | 0.00% | 132.7 |
| 1 | 0.1 | 0.1 | 5.33% | 0.33% | 3.37% | 90.97% | 9.1 |
| 0.3 | 0.1 | 0.1 | 6.62% | 0.41% | 3.35% | 89.62% | 9.0 |
| 0.1 | 0.1 | 0.1 | 7.15% | 0.43% | 3.35% | 89.07% | 8.9 |
| 0.03 | 0.1 | 0.1 | 7.32% | 0.44% | 3.34% | 88.90% | 8.9 |
| 1 | 1 | 0 | 69.20% | 5.06% | 25.74% | 0.00% | 56.3 |
| 0.3 | 1 | 0 | 80.04% | 5.15% | 14.81% | 0.00% | 32.1 |
| 0.1 | 1 | 0 | 83.35% | 5.14% | 11.51% | 0.00% | 24.6 |
| 0.03 | 1 | 0 | 84.64% | 5.16% | 10.20% | 0.00% | 21.8 |
| 1 | 1 | 0.1 | 10.66% | 0.77% | 3.88% | 84.69% | 8.5 |
| 0.3 | 1 | 0.1 | 19.86% | 1.27% | 3.47% | 75.40% | 7.5 |
| 0.1 | 1 | 0.1 | 25.50% | 1.56% | 3.27% | 69.68% | 7.0 |
| 0.03 | 1 | 0.1 | 28.32% | 1.72% | 3.15% | 66.81% | 6.7 |

4. Discussion

The purpose of this work was to understand how bare *n*-type GaAs crystals with external reflectors could have luminosities as high as 110 photons/keV at 10K [3], comparable to the brightest known scintillators. The high luminosity, low band gap, apparent absence of afterglow [1], and commercial availability as electronic-grade kg crystals make scintillating GaAs an attractive target for the detection of rare, low-energy electronic excitations from interacting dark matter particles.

The Monte Carlo results presented in this paper show that the observed high luminosity can be explained if (1) there is an optical scattering mechanism that allows photons to enter the escape cone of the exit face, and (2) the absolute absorption coefficient is below 0.1/cm. One possible explanation for the optical scattering in *n*-type GaAs is that scintillation photons are absorbed by the free electrons and promptly re-emitted with essentially the same energy, as in a metallic mirror. *N*-type GaAs above the Mott free-electron concentration limit of 8 x 10$^{15}$/cm$^3$ is metallic in character and electrically conductive, even at cryogenic temperatures [10]. The infrared narrow-beam absorption coefficient reported at 90K [6] is proportional to the free electron concentration, ranging from 0.1/cm for 10$^{16}$/cm$^3$ to 10/cm for 10$^{18}$/cm$^3$, which corresponds to a free electron cross section of 10$^{-17}$ cm$^2$. The absorption coefficient reported at 100 K [7] is 2.5/cm at 4.7 x 10$^{17}$/cm$^3$, which corresponds to a free electron cross section of 0.5 x 10$^{-17}$ cm$^2$. A 7.2 nm ultra-thin gold mirror (which also has free electrons bound to positive ions) reflects about 50%, transmits about 50%, and absolutely absorbs a much smaller fraction [11]. This corresponds to a free electron cross section of 1.7 x 10$^{-17}$ cm$^2$. This shows that free electrons in *n*-type GaAs and those in a thin gold mirror interact with photons with similar cross sections. An important difference is that free electrons in a gold mirror reflect photons in a preferred direction while free electrons in *n*-type GaAs should scatter photons randomly. First principles calculations should be able to determine the optical scattering cross section and angular distribution (assumed to be isotropic in this paper) as well as the absolute absorption cross section.

The exit fraction could be increased by using a photonic layer on the exit face [12]. This would increase the exit cone angle, and decrease the average path length and absolute absorption in the crystal.

5. Conclusions
- If the narrow-beam absorption coefficient in a 1 cm *n*-type GaAs cube is 1/cm and only due to absolute absorption, the exit fractions for the 16 surface treatments range from 1.1% to 2.1%.
- If the narrow-beam absorption coefficient in a 1 cm *n*-type GaAs cube is 1/cm and only due to optical scattering, the exit fractions for the 16 different surface treatments range from 30% to 69%. The 69% exit fractions are achieved with bare polished surfaces and external reflectors.
- The high scintillation luminosities reported at 10K for bare *n*-type GaAs crystals with external reflectors [1, 3] are consistent with scattering coefficients above 0.1/cm and absolute absorption coefficients below 0.1/cm.
- Future experimental measurements of internal scattering and absolute absorption (described in Appendix B) can provide new knowledge of the optical properties of *n*-type GaAs.

Acknowledgements

I thank M. Garcia-Sciveres, F. Moretti, and M. Pyle for helpful discussions. This work was supported by a U.S. Department of Energy Quantum Information Science Enabled Discovery (QuantISED) grant for High Energy Physics (KA2401032).

# Appendix A
# Steps in the Python Monte Carlo Code

The Monte Carlo calculations were performed by a custom Python code written by the author. The GaAs crystal is rectangular with dimensions $A_x$, $A_y$, and $A_z$, oriented to the Cartesian axes, and centered at 0,0,0. The crystal lies within six planes. The $z = +A_z/2$ plane contains the exit (top) face and the $z = -A_z/2$ plane contains the bottom face. The four side faces lie in the planes defined by $x = +A_x/2$, $x = -A_x/2$, $y = +A_y/2$, and $y = -A_y/2$. The exit face is modeled as a polished surface with an external 100% absorber. The other faces are polished with an optically bonded diffuse reflector (PD), a surface mirror (PM), an external diffuse reflector (PED), or an external mirror (PEM). The program steps are described below in sufficient detail for implementation in any scientific programming language:

1. Generate a photon from a uniformly random distribution of points within the crystal and from an isotropically random distribution in angle. For the latter, select three random trial direction cosines $d_x$, $d_y$, and $d_z$, each from a uniform distribution between −1. and +1. Choose the first set that satisfies the condition $D^2 = d_x^2 + d_y^2 + d_z^2 <= 1$. Scale to a unit vector by dividing each trial cosine by $D$. If the photon travels from point $x, y, z$ along path length $L$, the end point coordinates are $x + L(d_x/D)$, $y + L(d_y/D)$, and $z + L(d_z/D)$.
2. Use the position and scaled direction cosines from step 1 to find the shortest path length $L_F$ that intersects a crystal face plane. Select two random numbers $R_S$ and $R_A$ from a uniform distribution between 0 and 1. Compute an exponentially distributed random scatter path length $L_S$ and absorption path length $L_A$ using the optical scattering coefficient $B_S$ and optical absorption coefficient $B_A$: $L_S = -\ln(R_S)/B_S$ and $L_A = -\ln(R_A)/B_A$,
2.1 If $L_A$ is shorter than both $L_S$ and $L_F$, tally the photon as absorbed and go to step 1 to generate a new photon.
2.2 If $L_S$ is shorter than both $L_A$ and $L_F$, the photon has scattered in the crystal. Use $L_S$ to compute the interaction position. Select a new set of isotropically random direction cosines as in step 1 and go to step 2. to continue tracking the photon.
2.3 If $L_F$ is shorter than both $L_S$ and $L_A$, use the intersection point at the face determined in step 2 as the new position. If the face is PD go to step 3. If the face is PM, go to step 4. If the face is PED or PEM, go to step 5.
3. Case PD surface. The face is polished and optically bonded to a diffuse reflector. There is no Fresnel reflection or transmission. Select a random number from a uniform distribution between 0 and 1.
3.1 If the random number is greater than the diffuse reflectance, tally the photon as absorbed and go to step 1 to generate a new photon.

3.2 If the random number is less than the diffuse reflectance, generate new random direction cosines that are isotropically distributed in the hemisphere that extends into the crystal. Go to step 2 to continue tracking the photon.

4. Case PM surface. The face is polished and coated with a mirror reflector. There is no Fresnel reflection or transmission. Select a random number from a uniform distribution between 0 and 1.

4.1 If the random number is greater than the mirror reflectance, tally the photon as absorbed and go to step 1 to generate a new photon.

4.2 If the random number is less than the mirror reflectance, the photon reflects back into the crystal. Change the sign of the direction cosine that is parallel to the face normal. Go to step 2 to continue tracking the photon.

5. Case PED or PEM surface. The face is polished and has an external diffuse or mirror reflector. When the photon reaches the inner surface, it will either be Fresnel reflected back into the crystal or refracted out of the crystal ($n_1 = 3.5$) into the narrow space between the crystal and the external reflector ($n_2 = 1$). Use the direction cosine parallel to the face normal to compute the angle of incidence $\theta$.

5.1 If $\sin(\theta) \geq 1/n_1$, the photon is totally internally reflected back into the crystal. Change the sign of the direction cosine that is parallel to the face normal. Go to step 2 to continue tracking the photon.

*Note*: A photon is trapped by total internal reflection when the absolute values of all three direction cosines are less than the critical cosine for total internal reflection. For GaAs, the critical $\sin(\theta)$ is $1/3.5 = 0.286$ and the critical $\cos(\theta)$ is 0.958. At each reflection the direction cosine parallel to the face normal changes sign but the absolute value does not change.

5.2 If $\sin(\theta) < 1/n_1$ the photon is in the escape cone and the Fresnel reflectance is given by:

$$R(\theta, n_1, n_2) = 0.5 \left( \frac{n_1 \cos(\theta) - n_2 \sqrt{1 - \left(\frac{n_1}{n_2} \sin(\theta)\right)^2}}{n_1 \cos(\theta) + n_2 \sqrt{1 - \left(\frac{n_1}{n_2} \sin(\theta)\right)^2}} \right)^2 + 0.5 \left( \frac{n_1 \sqrt{1 - \left(\frac{n_1}{n_2} \sin(\theta)\right)^2} - n_2 \cos(\theta)}{n_1 \sqrt{1 - \left(\frac{n_1}{n_2} \sin(\theta)\right)^2} + n_2 \cos(\theta)} \right)^2$$

Select a random number from a uniform distribution between 0 and 1.

5.3 If the random number is less than the Fresnel reflectance, the photon is reflected back into the crystal. Change the sign of the direction cosine that is parallel to the face normal. Go to step 2 to continue tracking the photon.

5.4 If the random number is greater than the Fresnel reflectance, the photon is transmitted out of the crystal. If the face is the exit face, tally the photon as absorbed (detected) and go to step 1 to generate a new photon. Otherwise, the photon will reach the external reflector in the PED or PEM cases below. Refract the photon from internal angle $\theta_1$ to external angle $\theta_2$ by Snell's law: $n_1 \sin(\theta_1) = n_2 \sin(\theta_2)$. The new direction cosine parallel to the face normal is $\cos(\theta_2)$. Since the optical momentum in the other two directions is conserved, multiply those direction cosines by $n_1/n_2$. The result is a set of unit vector direction cosines in the narrow space between the crystal and the external reflector.

*Note*: This calculation of direction cosines is not necessary for case PED, since they will be replaced with values from a random hemisphere in step 6.2, but has been included in the spirit of tracking each photon at every surface interaction.

6. Case PED. The photon exits the crystal (step 5.4) into the narrow gap between the crystal and the external diffuse reflector. Select a random number from a uniform distribution between 0 and 1.

6.1 If the random number is greater than the diffuse reflectance, tally the photon as absorbed. Go to step 1 to generate a new photon.

6.2 If the random number is less than the diffuse reflectance, the photon is reflected toward the crystal in the narrow space between the diffuse reflector and the crystal. Generate new random direction cosines that are isotropically distributed in the hemisphere that faces the crystal surface. Use the direction cosine parallel to the face normal to compute the angle $\theta_1$. Compute the Fresnel reflectance $R(\theta_1, n_1, n_2)$, where $n_1 = 1$ and $n_2 = 3.5$. Select a random number from a uniform distribution between 0 and 1.

6.2.1 If the random number is greater than $R(\theta_1, n_1, n_2)$, the photon enters the crystal and is refracted from external angle $\theta_1$ to internal angle $\theta_2$ by Snell's law: $n_1 \sin(\theta_1) = n_2 \sin(\theta_2)$. Compute the new direction cosine parallel to the face normal as $\cos(\theta_2)$. Since the optical momentum in the other two directions is conserved, multiply those direction cosines by $n_1/n_2$. The result is a set of unit vector direction cosines in the crystal. Go to step 2 to continue tracking the photon inside the crystal.

6.2.2 If the random number is less than $R(\theta_1, n_1, n_2)$, the photon is reflected from the crystal back toward the external reflector. Go back to step 6 and continue cycling until the photon is either absorbed by the external diffuse reflector in step 6.1 or until it enters the crystal in step 6.2.1.

7. Case PEM. The photon exits the crystal (step 5.4) into the narrow gap between the crystal and the external mirror. Select a random number from a uniform distribution between 0 and 1.

7.1 If the random number is greater than the mirror reflectance, tally the photon as absorbed. Go to step 1 to generate a new photon.

7.2 If the random number is less than the mirror reflectance, the photon is reflected back toward the crystal in the narrow space between the mirror and the crystal. Change the sign of the direction cosine that is parallel to the face normal. Compute the Fresnel reflectance $R(\theta_1, n_1, n_2)$, where $n_1 = 1$ and $n_2 = 3.5$. Select a random number from a uniform distribution between 0 and 1.

7.2.1 If the random number is greater than $R(\theta_1, n_1, n_2)$, the photon enters the crystal and is refracted from external angle $\theta_1$ to internal angle $\theta_2$ by Snell's law: $n_1 \sin(\theta_1) = n_2 \sin(\theta_2)$. Compute the new direction cosine parallel to the face normal as $\cos(\theta_2)$. Since the optical momentum in the other two directions is conserved, multiply those direction cosines by $n_1/n_2$. The result is a set of unit vector direction cosines in the crystal. Go to step 2 to continue tracking the photon inside the crystal.

7.2.2 If the random number is less than $R(\theta_1, n_1, n_2)$, the photon is reflected from the crystal back toward the external mirror. Go back to step 7 and continue cycling until the photon is either absorbed by the external mirror in step 7.1 or enters the crystal in step 7.2.1.

## Appendix B
## Methods for the Experimental Measurement of Absolute Absorption and Scatter.

The GaAs crystal should be well polished and free from any surface scatter or absorption.

*Step 1. Narrow beam experiment*
The fraction of the narrow beam that is transmitted through the sample is given by

$$f_T = \sum_{n=0}^{\infty} (1-r)^2 t (r^2 t^2)^n = \frac{(1-r)^2 t}{1 - r^2 t^2}$$

The single surface reflectance is $r = (1-n)^2/(1+n)^2$, where $n$ is the refractive index of GaAs, and $t$ is the narrow-beam transmission through sample thickness $d$. Solving for $t$:

$$t = \frac{\sqrt{(1-r)^4 + 4f_T^2 r^2} - (1-r)^2}{2 f_T r^2}$$

The fraction of the narrow beam that is reflected by the sample is given by

$$f_R = r + \sum_{n=0}^{\infty} (1-r)^2 r t^2 (r^2 t^2)^n = r + \frac{(1-r)^2 r t^2}{1 - r^2 t^2}$$

The sum of the fractions of scattered and absolutely absorbed narrow beam photons is

$$f_A + f_S = 1 - f_R - f_T.$$

*Step 2. Integrating sphere experiment.*

When the integrating sphere is empty the detector will measure $C_0$ photons, where $C_0 = N (M_0 \, A_D/A_S)$ and $N$ is the number of narrow-beam photons entering the sphere. $M_0$ is the empty sphere multiplier, and $A_D$ and $A_S$ are the detector and sphere areas. In this case the narrow beam strikes the opposite wall and scatters until it is absolutely absorbed by the walls, the sample holder, the sample, or the detector. The sphere multiplier is effectively the number of times that the photons scatter before they are absolutely absorbed.

When the sample is in the sphere and swung away from the beam the detector will measure $C_{away}$ photons where $C_{away} = N (M_{away} \, A_D/A_S)$. $M_{away}$ will be slightly lower than $M_0$ due to sample and holder absorption of the circulating photons.

When the sample is in the beam the detector will measure $C_{in}$ photons, where

$C_{in} = N (f_R + f_T + f_S)(M_{in} \, A_D/A_S)$

It is reasonable to assume that $M_{in} = M_{away}$, since the sample and holder should absolutely absorb the same fraction of the circulating beam whether it is in the beam or swung away from the beam.

$C_{in}/C_{away} = f_R + f_T + f_S = 1 - f_A$

The fraction absolutely absorbed by sample is $f_A = 1 - C_{in}/C_{away}$. The integrating sphere determines the absolute absorption and is not sensitive to the scattering that the narrow beam experiment detects